\documentclass[a4paper,11pt]{article}

\pdfoutput=1 
\usepackage{jinstpub} 

\title{\boldmath The Construction of the Fiber-SiPM beam monitor system 
of the R484 and R582 Experiments at the RIKEN-RAL Muon facility.}

\author[a,1]{M.Bonesini,\note{Corresponding author.}}
\author[a]{R.Bertoni,}
\author[a]{F.Chignoli,}
\author[a]{R.Mazza,}
\author[b]{T.Cervi,}
\author[b]{A.deBari,}
\author[b]{A.Menegolli,}
\author[b]{M.C.Prata,}
\author[b]{M.Rossella,}
\author[c]{L.Tortora,}
\author[d]{R.Carbone,}
\author[d]{E.Mocchiutti,}
\author[d,2]{A.Vacchi,\note{Also at Department of Mathematics, Computer Science
and Physics, University of Udine, Udine, Italy.}}
\author[d]{E.Vallazza,}
\author[d]{G.Zampa}

\affiliation[a]{Sezione INFN Milano Bicocca, Dipartimento di Fisica 
G. Occhialini, Universit\'a di Milano Bicocca, \\
Piazza Scienza 3, Milano, Italy}
\affiliation[b]{Sezione INFN Pavia, Dipartimento di Fisica, 
 Universit\'a di Pavia, \\ via A. Bassi 6, Pavia, Italy}
\affiliation[c]{Sezione INFN Roma Tre, \\ via della Vasca Navale 84, Roma, Italy}
\affiliation[d]{Sezione INFN Trieste, \\ via Padriciano 99, Trieste, Italy}

\emailAdd{maurizio.bonesini@mib.infn.it}

\abstract{The scintillating fiber-SiPM beam monitor detectors, 
designed to deliver beam informations for the R484 and R582
experiments at the high intensity, low energy
pulsed muon beam at the RIKEN-RAL facility, have 
been successfully constructed and operated. Details on their construction and first performances in beam are reported. }


\begin{document}
\maketitle
\flushbottom

\section{Introduction}
\label{sec:intro}
The FAMU experiment at RAL, see reference \cite{famu} for further details,
aims at the measurement of the hyperfine splitting (HFS) in the ground
state (1S) of the muonic hydrogen \cite{vacchi}, thus providing a high accuracy 
determination of the proton Zemach radius \cite{zemach}. 
The experiment may thus contribute to solve the so-called ``proton 
radius puzzle'': a 7 $\sigma$ disagreement between the proton charge
radius as determined from electrons or muons \cite{pohl}. 
\begin{figure}[htbp]
\centering 
\includegraphics[width=.45\textwidth]{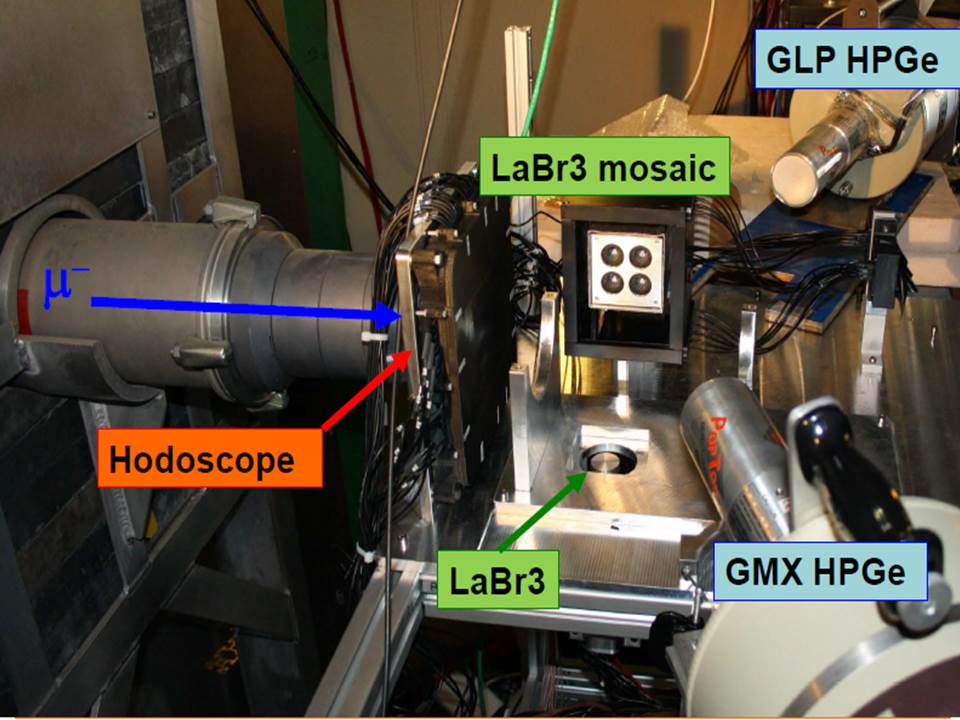}
\includegraphics[width=.50\textwidth]{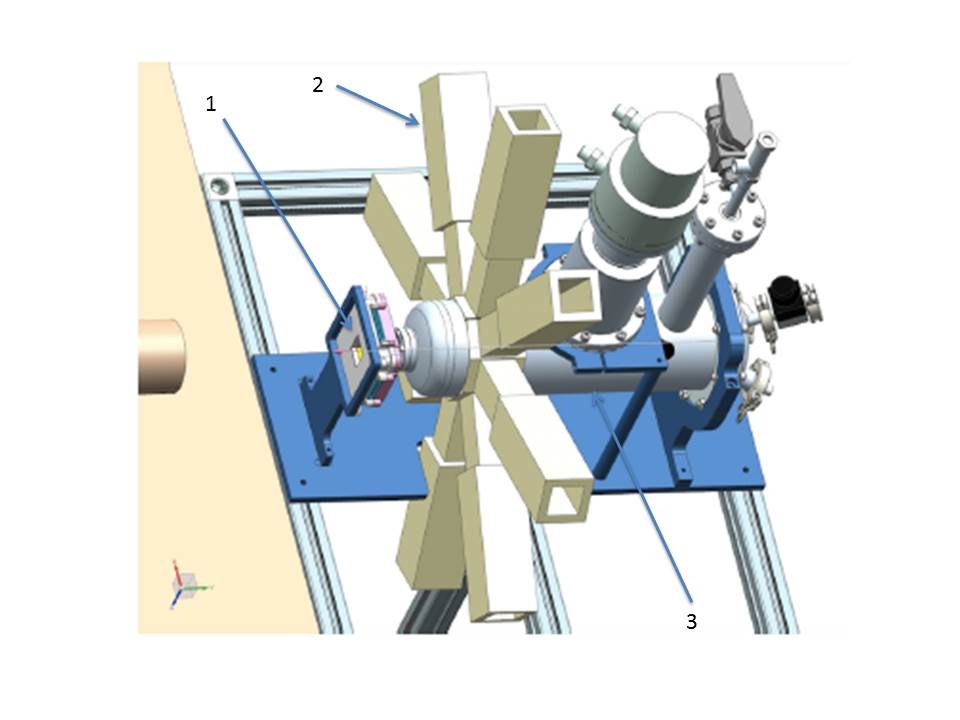}
\caption{Left panel: picture of the setup for the 2014 data-taking (R484).
Right panel: layout of the setup for the 2015-2016 data-taking (R582); 1)
is the beam hodoscope, 2) the crown of 8 LaBr$_3$ crystals read by 
photomultipliers  and 3) the
cryogenic target. The four HPGe detectors, also used in this run, are not 
shown.  }
\label{fig1}
\end{figure}
For this experiment, an important issue is the optimal steering of the
impinging high intensity pulsed muon beam onto the hydrogen target, to
maximize the muonic hydrogen production rate.
A system of two beam hodoscopes has been developed for this scope. 
The first one is based on square $3 \times 3$ mm$^2$ scintillating fibers read 
by SiPM, for the R484 experiment, which had a data taking in June 2014, while
the second one is based on square $1 \times 1$ mm$^2$ scintillating fibers
read by SiPM, for the R582 experiment, which instead had a data taking in
December 2015 and February 2016. 
A schematic layout of both experimental setup is shown in figure \ref{fig1}.
  
\begin{figure}[htbp]
\centering 
\includegraphics[width=.5\textwidth]{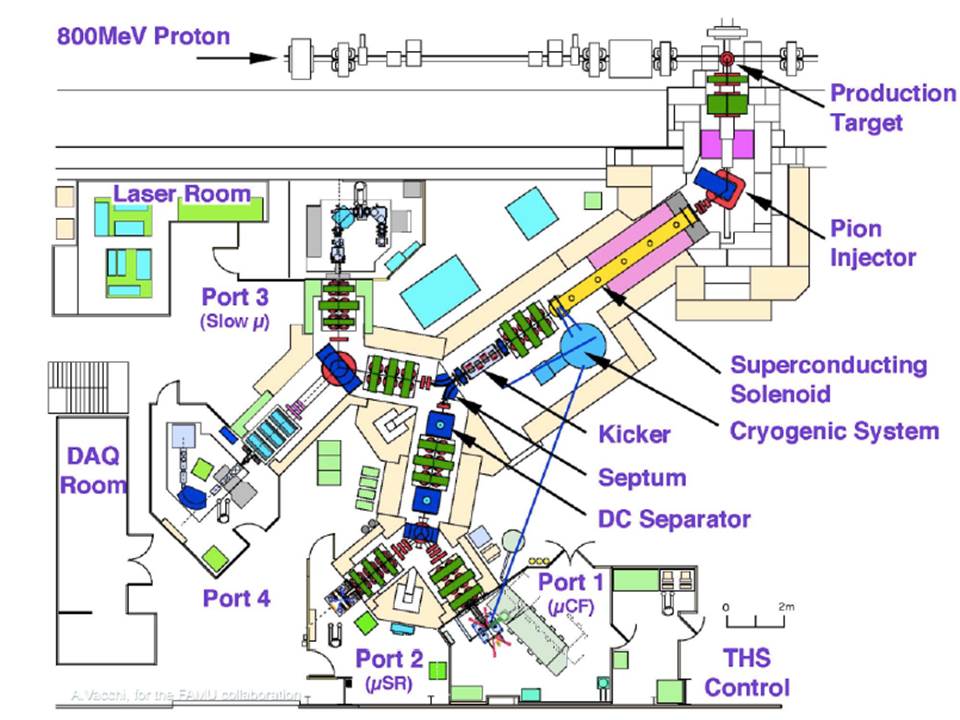}
\includegraphics[width=.49\textwidth]{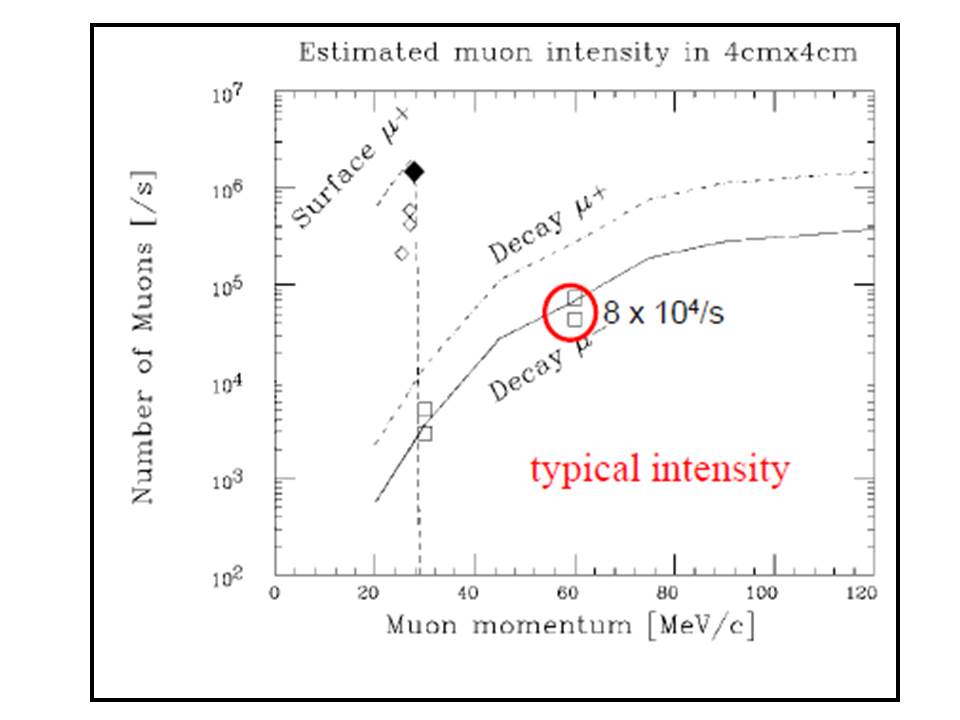}
\caption{Left panel: schematic layout of the RIKEN-RAL muon facility at RAL, 
with its four experimental ports. The FAMU experiment presently uses port 4 
and will move soon to port 1 for its final run. Right panel: estimated 
intensity in a $4 \times 4$ cm$^2$ area for surface and decay muon beams
at RIKEN-RAL.}
\label{fig2}
\end{figure}

The RIKEN-RAL muon facility at Rutherford Appleton Laboratory (UK) provides
high intensity pulsed muon beams at four experimental ports, as shown in
figure \ref{fig2}. 
The ISIS primary proton beam at 800 MeV/c, with a 50 Hz repetition rate, 
impinges from the left on a secondary carbon target producing pions and then
high intensity
low energy pulsed muon beams. Surface $\mu^{+}$ (20-30 MeV/c) are produced
by pions stopping close to the target surface, while  
decay $\mu^{+}/\mu^{-}$ (20-120 MeV/c) are produced from pion decay
outside the target. 

The muon beams reflect the primary beam structure: two pulses with a 70 ns 
FWHM and a 320 ns peak to peak distance are delivered. The FAMU experiment
makes use of a negative 
decay muon beam at $\sim$ 60 MeV/c. The intensity is around 
$8 \times 10^{4} \ \mu^{-}$/s in a typical size $4 \times 4$ cm$^2$ 
 as shown in the  right  panel of figure \ref{fig2}.
The energy spread is around $10 \ \%$ and the angular divergence around 
60 mrad. 

\section{Construction of the R484 and R582 beam hodoscopes}
\label{sec:constr}
At the entrance of the FAMU target, the beam profile is nearly flat (see later)
and thus we may expect $\sim 20 \ \mu /$ mm in the 70 ns long spill (along 
transverse x/y directions). This fact puts severe constraints on the beam hodoscope
construction and prevents a particle by particle identification. 
\subsection{The R484 experiment beam hodoscope}
\begin{figure}[htbp]
\centering 
\includegraphics[width=.49\textwidth]{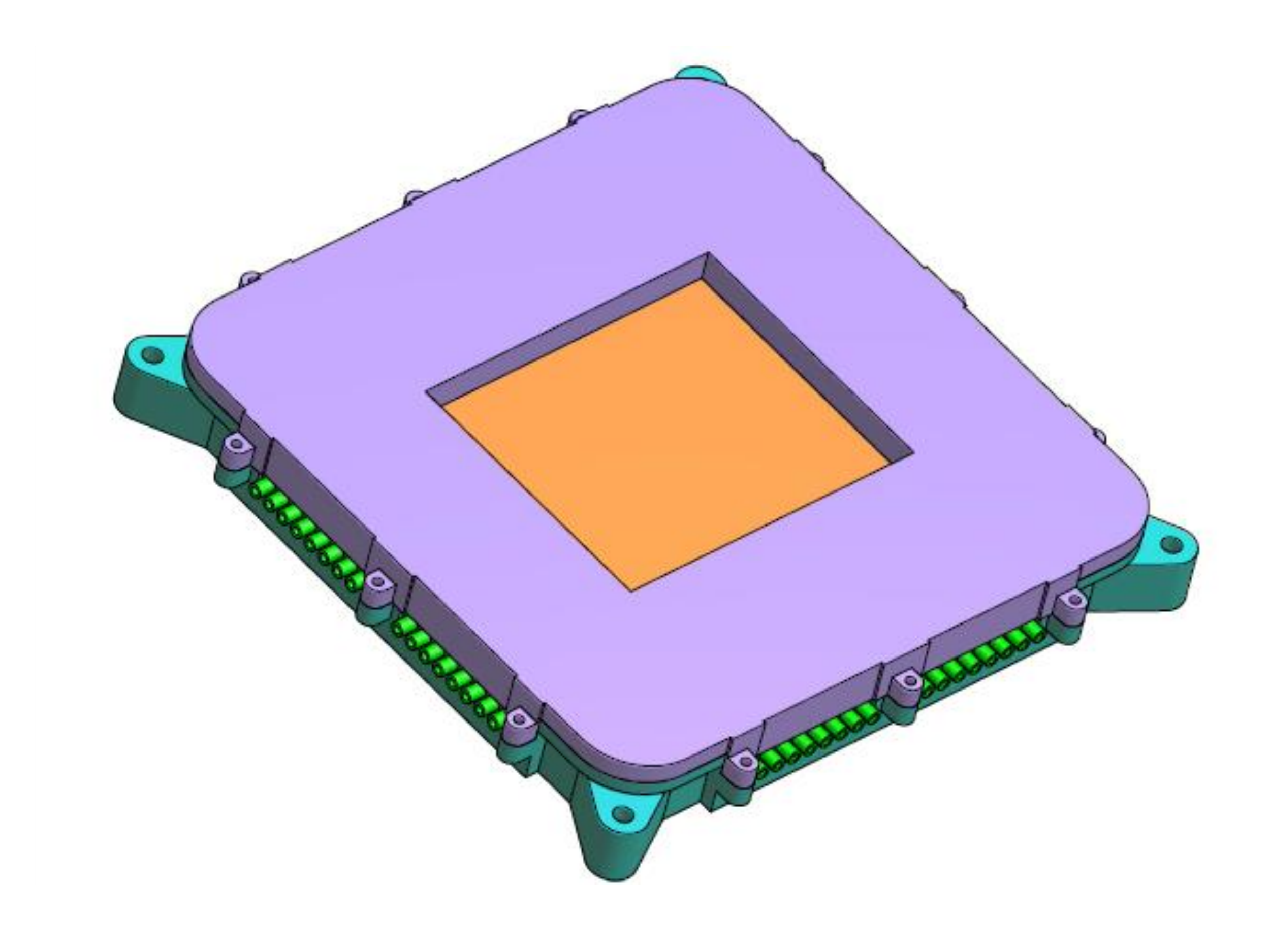}
\includegraphics[width=.49\textwidth]{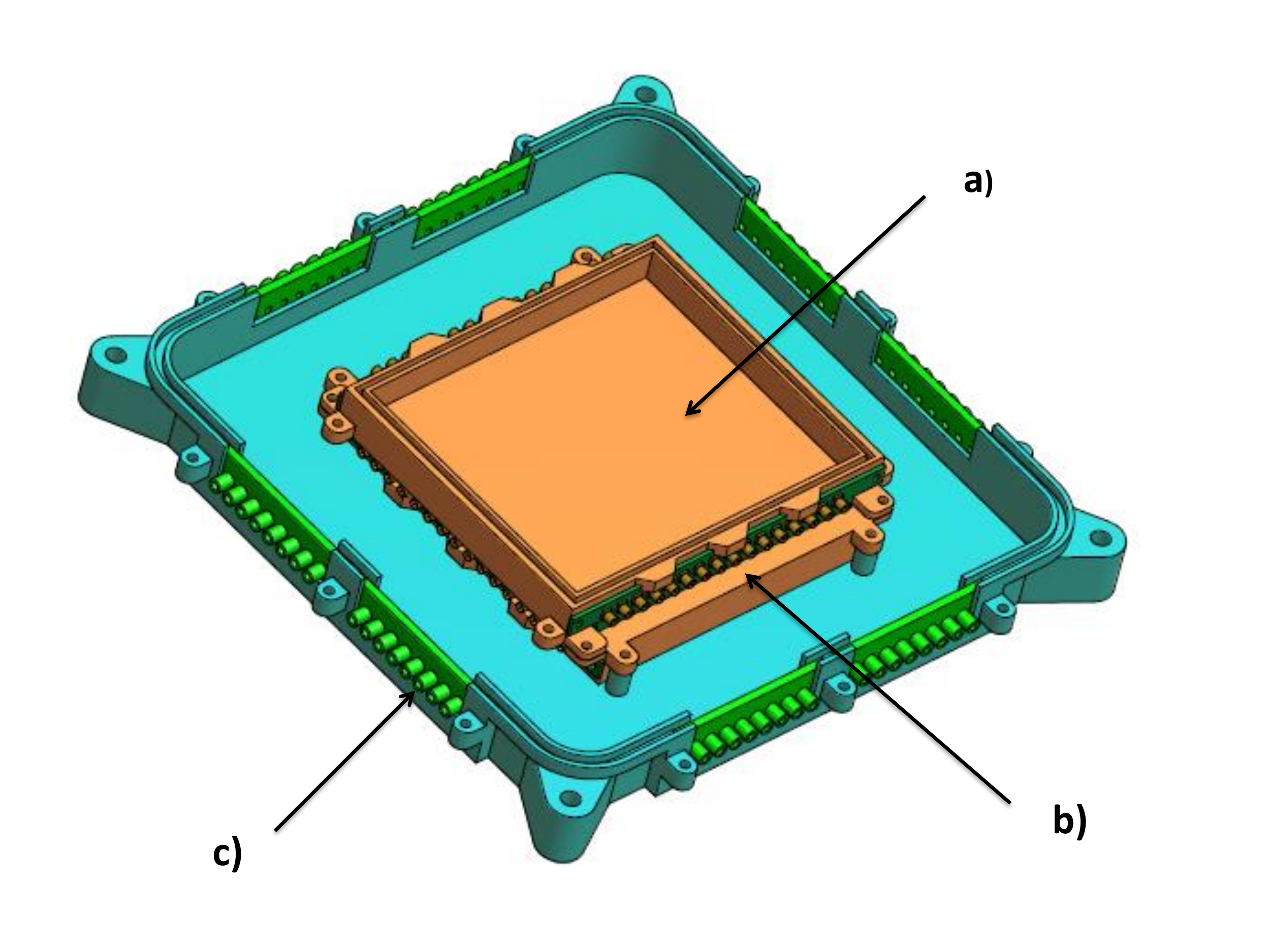}
\caption{Mechanics layout of the R484 3 mm pitch beam hodoscope. 
Left panel: general structure of the light-tight assembly. Right panel:
details of one of the two cover boxes.  a) holder
for the scintillating fibers; b) custom PCB with SiPMs on one side 
(towards the fiber face) and MCX connectors, to convey signal and power
for each individual channel,  on the other side; 
c) cable feed-through on the lateral side of the structure.}
\label{fig-ass}
\end{figure}

\begin{figure}[htbp]
\centering 
\vskip -5cm
\includegraphics[width=.35\textwidth]{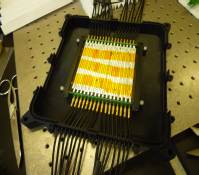}
\includegraphics[width=.5\textwidth]{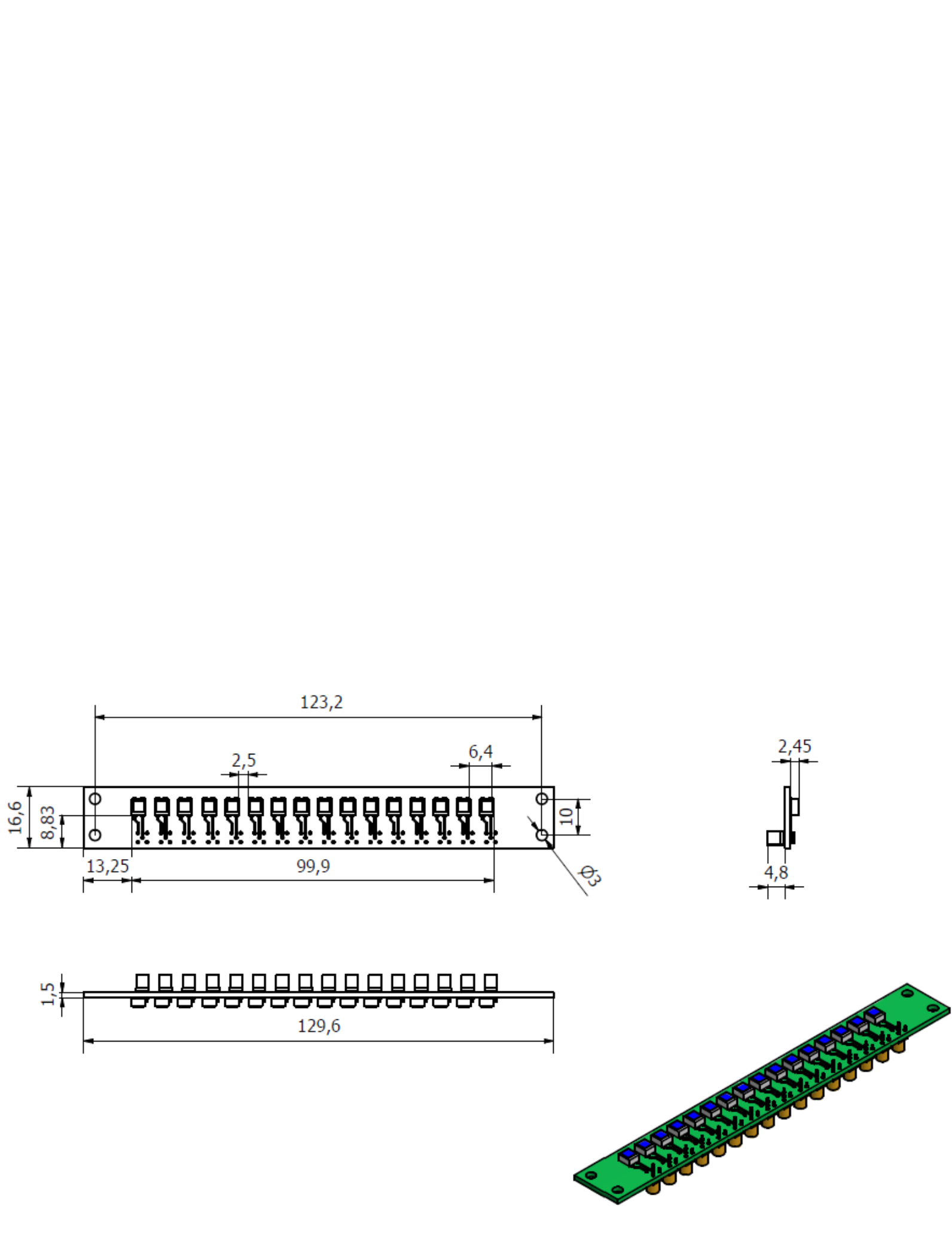}
\caption{Left panel: hodoscope assembly: one layer of fiber is visible, inside the fiber
holder. Each fiber is facing a SiPM mounted on one side of a custom PCB. All is
already housed inside one of the two ABS cover boxes. RG174 cables, with 
MCX connectors on one side, are visible and extend beyond the cover box 
volume, through feedthrougs. Right panel: schematic layout of the custom 
PCB holding 
on the top side the 16 SiPMs, facing fiber edges and on the bottom 
side MCX connectors.}
\label{fig:pcb}
\end{figure}

\begin{figure}[htbp]
\centering 
\includegraphics[width=.49\textwidth]{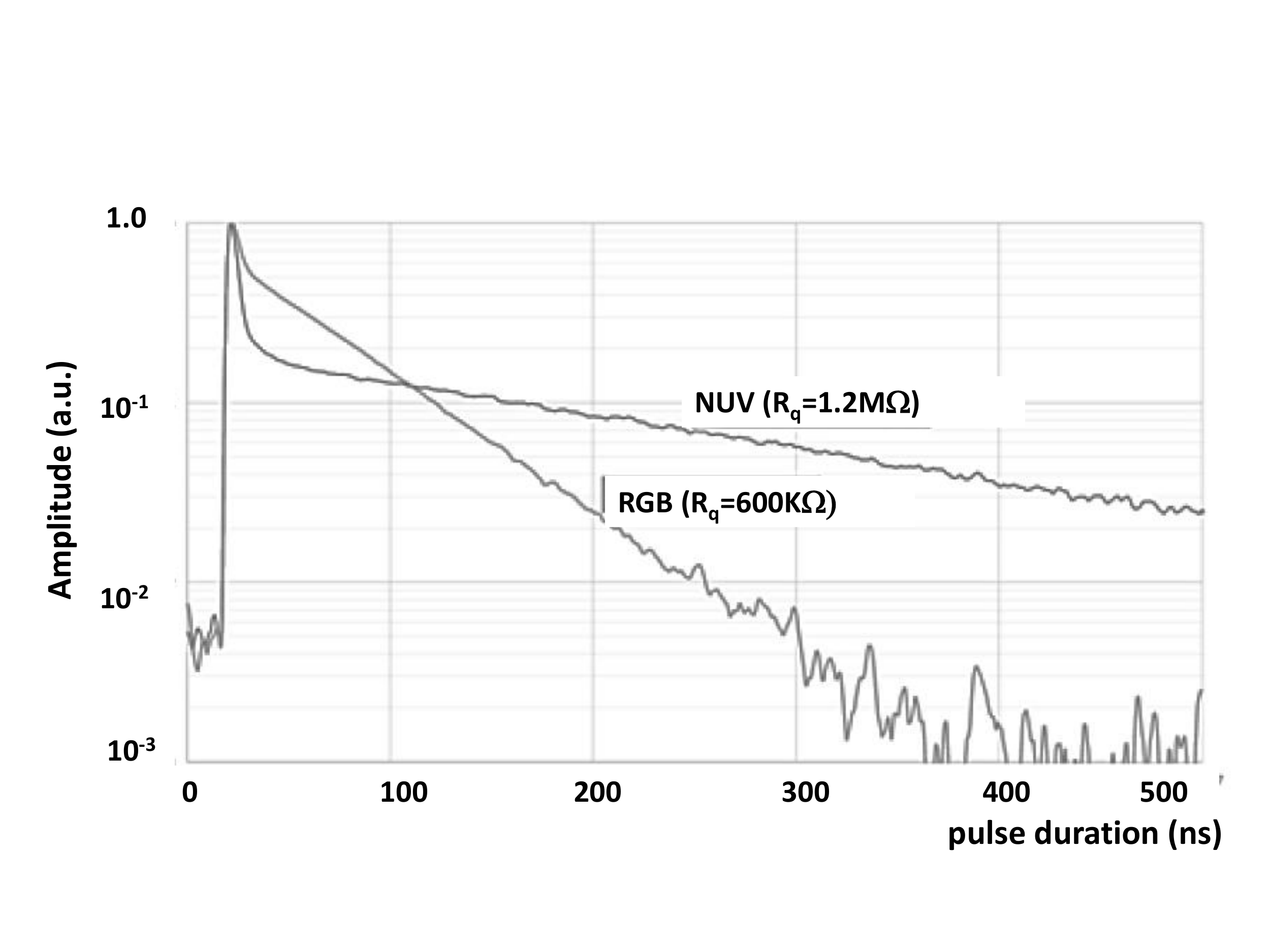}
\includegraphics[width=.49\textwidth]{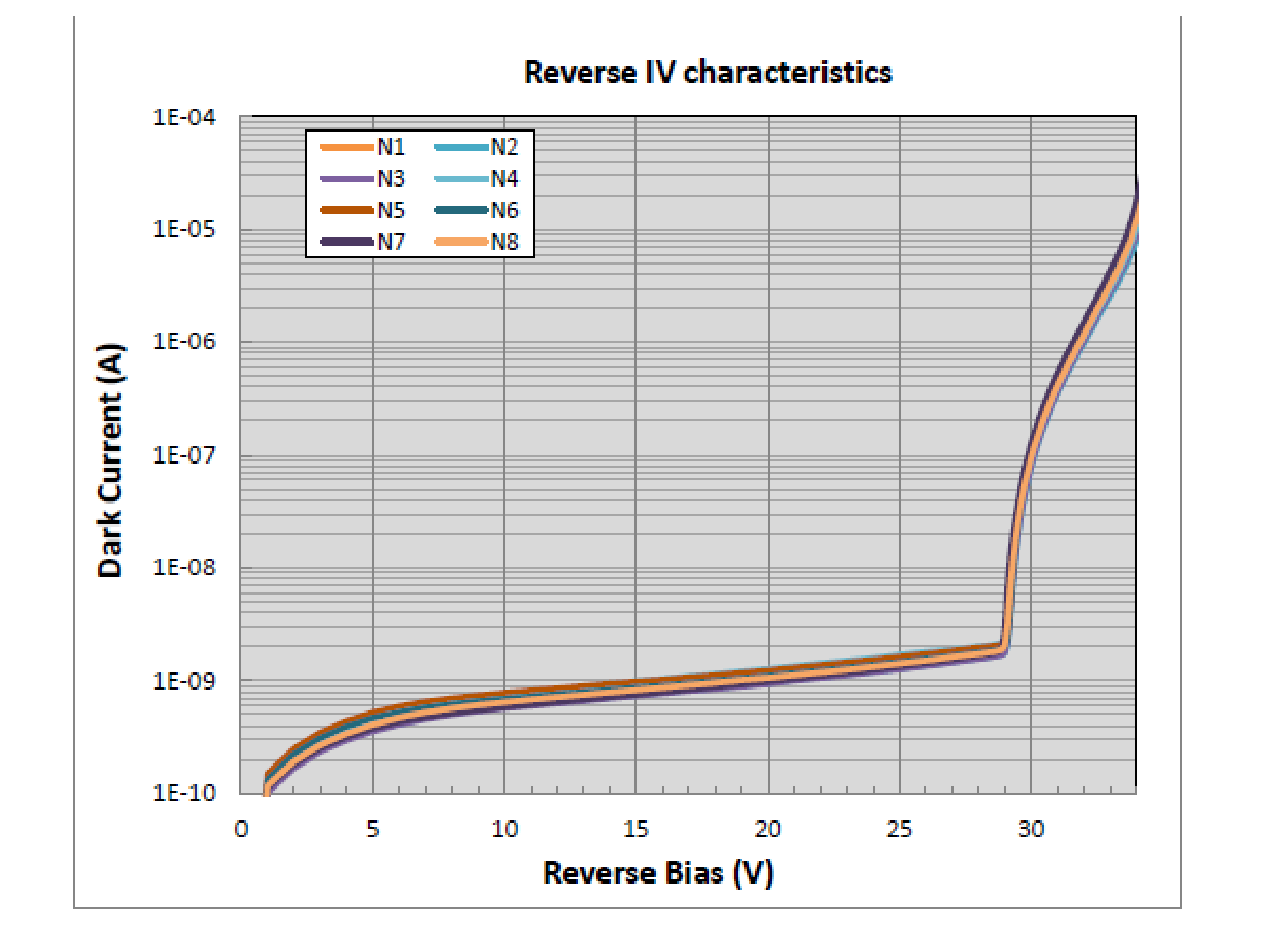}
\caption{Left panel: pulse duration for RGB and NUV types of
Advansid $3 \times 3 $ $mm^2$ SiPMs. The shorter pulse duration of the
RGB type ($\sim 200$ ns) has to be noticed [courtesy of Advansid srl]. 
Right panel:  reverse current-voltage
 characteristic 
for a sample of Advansid RGB SiPMs. 
The breakdown voltage is estimated
around 29.1 V. Measurements were done at $T_A=25 ^{\circ}C$. }
\label{fig:sipmt}
\end{figure}

The R484 beam hodoscope (see figures \ref{fig-ass} and \ref{fig:pcb} for
details) consists of 32+32 scintillating fibers with square cross-section
\footnote{Bicron BCF12 single clad, side 3 mm, peak emission $\sim 435$ nm,
trapping efficiency $\sim 7 \%$, attenuation length 270 cm, decay time 
$\sim 3.2$ ns, light yield $\sim 8000$ photons/MeV} arranged along X/Y axis
(orthogonal axis perpendicular to the beamline). 
The use of square fibers makes the detector response independent from the
position of the muon trajectory inside a fiber and minimizes the amount 
of dead spaces. 
The fibers have been cut
to a length of $\sim 102.5$ mm with a Fiberfin4 machine at CERN, 
that provides directly polished
ends ready to use, suited for a detector with a $10 \times 10$ cm$^2$ fiducial
area. The fiber thickness of 3+3 mm corresponds approximately to $\sim 7 \%$ 
of the range of a 60 MeV/c muon. 
A muon passing through a fiber produces scintillation light that is detected
at one end of the fiber by a $3 \times 3$ cm$^2$ Silicon photomultiplier
(SiPM)~\footnote{Advansid RGB type, with $40 \ \mu$m cells}. 
For space problems, as the SiPM package footprint is bigger than the fiber size,
fibers are read from one edge only, alternating left/right (fibers along 
X-axis) and up/down (fibers along Y-axis) sides.
Fibers were wrapped with an Al-film, $10 \mu$m  thick, to avoid channel to
channel light cross-talk. 
Single RG174 cables, with MCX connectors on one side and LEMO00 connectors
on the other side, convey both the SiPM's bias (on the external braid 
shield) and signal ouput (on the cable inner connector). In this way, each 
SiPM may be powered individually.
Each SiPM is mounted on a custom printed board (in groups of 16), as shown
in figure \ref{fig:pcb}. SiPMs were hand-soldered on one side of the PCB, 
facing 
the scintillating fibers, while MCX connectors were mounted on the other
side of the PCB. 
The detector is housed in a 3D printed ABS case~\footnote{printed on a 
Stratasys Elite Dimension printer, with 0.18 mm resolution}.   
The choice of Advansid RGB SiPMs~\footnote{ASD-SiPM3S-P40, in a 
production batch, with a 
  $\sim 600K \Omega $ quenching resistor} as photodetectors for the R484 beam
hodoscope was dictated by their short pulse duration, their 
high photon efficiency, well matched to
the BC12 fiber peak emission (PDE $\sim 22 \%$ at $\sim 440$ nm, with 4 V 
overvoltage), their low
operating voltage ($V_{brk} \sim 29$V), their small breakdown voltage
dependence from temperature ($\sim 27 mV/C$)  and their low dark noise, 
see figure \ref{fig:sipmt} for details.
All $3 \times 3$ mm$^2$ SiPMs were individually charaterized to determine 
their breakdown voltage by measuring their current-voltage charateristic,
as shown in the right panel of figure \ref{fig:sipmt}.

The detector front-end was based on refurbished electronic boards from the
INFN TPS project \cite{tps}, that provide power to single SiPMs (up to
40 V). This feature allowed to use only  Advansid or SENSL SiPMs, 
excluding Hamamatsu or KETEC ones, that needed higher voltages. 
The TPS electronic system is based on a VME-like crate, providing power 
to custom boards and a GPIB module interface. Each TPS electronics board,
with 8 individual channels, provided individual channel voltage fine 
regulation, signal amplification and shaping, signal discrimination 
and trigger capability, by using the OR of the eight channels of a board.
Output signals are also  fed into a CAEN V792 QDC for measurement of the
charge integrated signal.

The beam hodoscope has been first tested with cosmics at INFN Milano Bicocca, 
where it has been  built,
 and at the Beam Test Facility (BTF) of the INFN LNF laboratory
\cite{btf}, before the R484 data-taking in mid 2014 at RAL. 
A custom data-acquisition system 
was based at first on the Bit3 VME-PCI interface and
afterwards on the  CAEN V2718 VME-PCI interface. 
Test beam results at BTF with impinging electrons~\footnote{490 MeV kinetic 
energy, nominal beam profile $9 \times 7$ mm$^2$, beam multiplicity 
5 particles/spill}, already reported in reference \cite{carbone}, show 
a signal to noise ratio better than 10 and a single to double MIP separation
at $\sim 1.5 \sigma$.  
With 60 MeV muons, we expect $\sim 300$ photons/muon arriving at the 
$3 \times 3$ mm$^2$ window of a SiPM, giving a collected charge of 
$\sim 7-14 $ pC at the QADC input, for a nominal SiPM's gain of about 
$0.8-1.5 \times 10^6$. In order to match the V792 QADC range (0-400 pC), 
with the
foreseen muon rate at RIKEN-RAL and avoid saturation, 10x attenuators 
had to be used before the QDC input.  
\subsection{The R582 experiment beam hodoscope}
In the following R582 experiment at RIKEN-RAL, a 1 mm pitch beam hodoscope,
to be put in front of the new cryogenic target, was developed. 
\begin{figure}[htbp]
\centering 
\includegraphics[width=.72\textwidth]{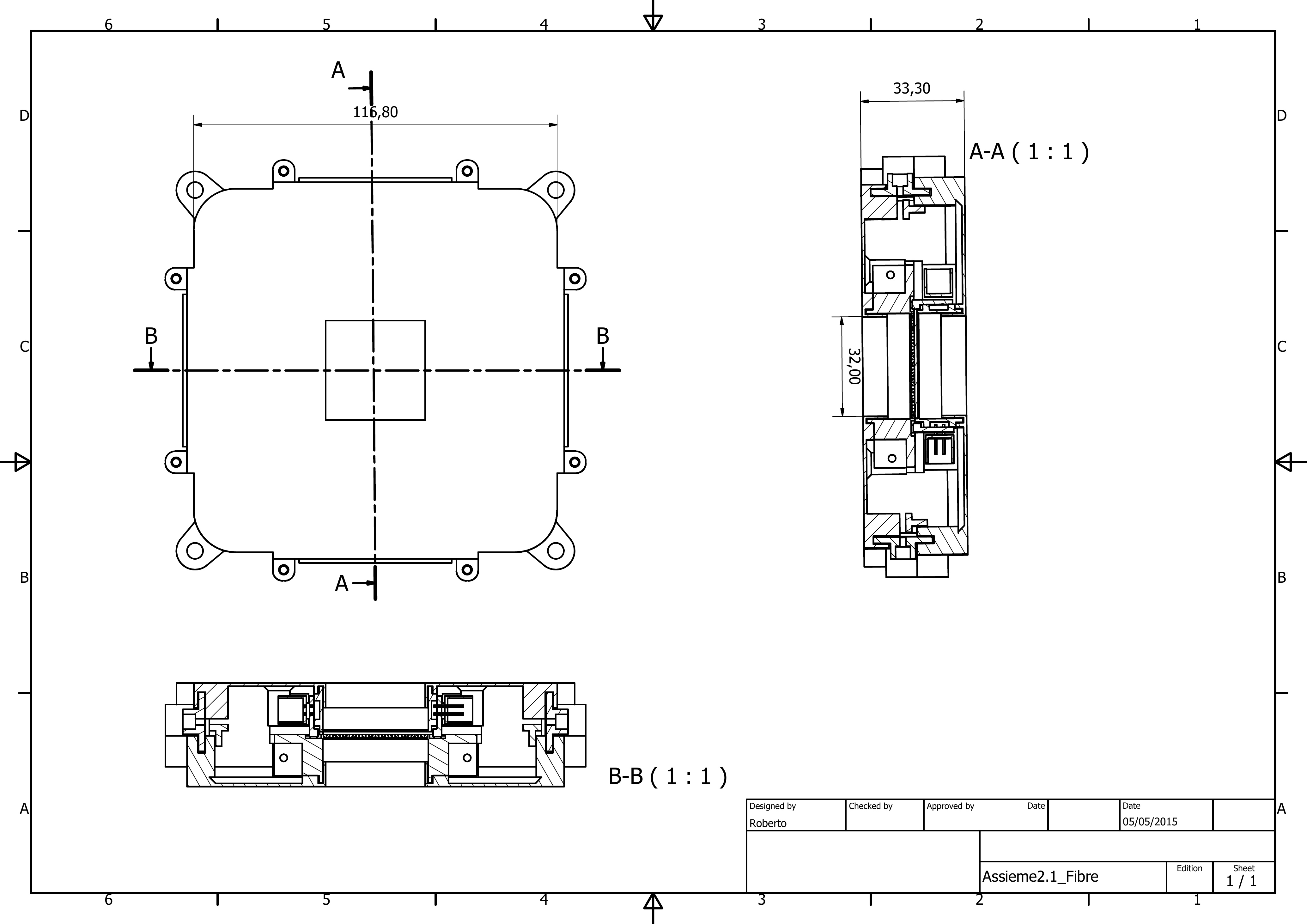}
\includegraphics[width=.59\textwidth]{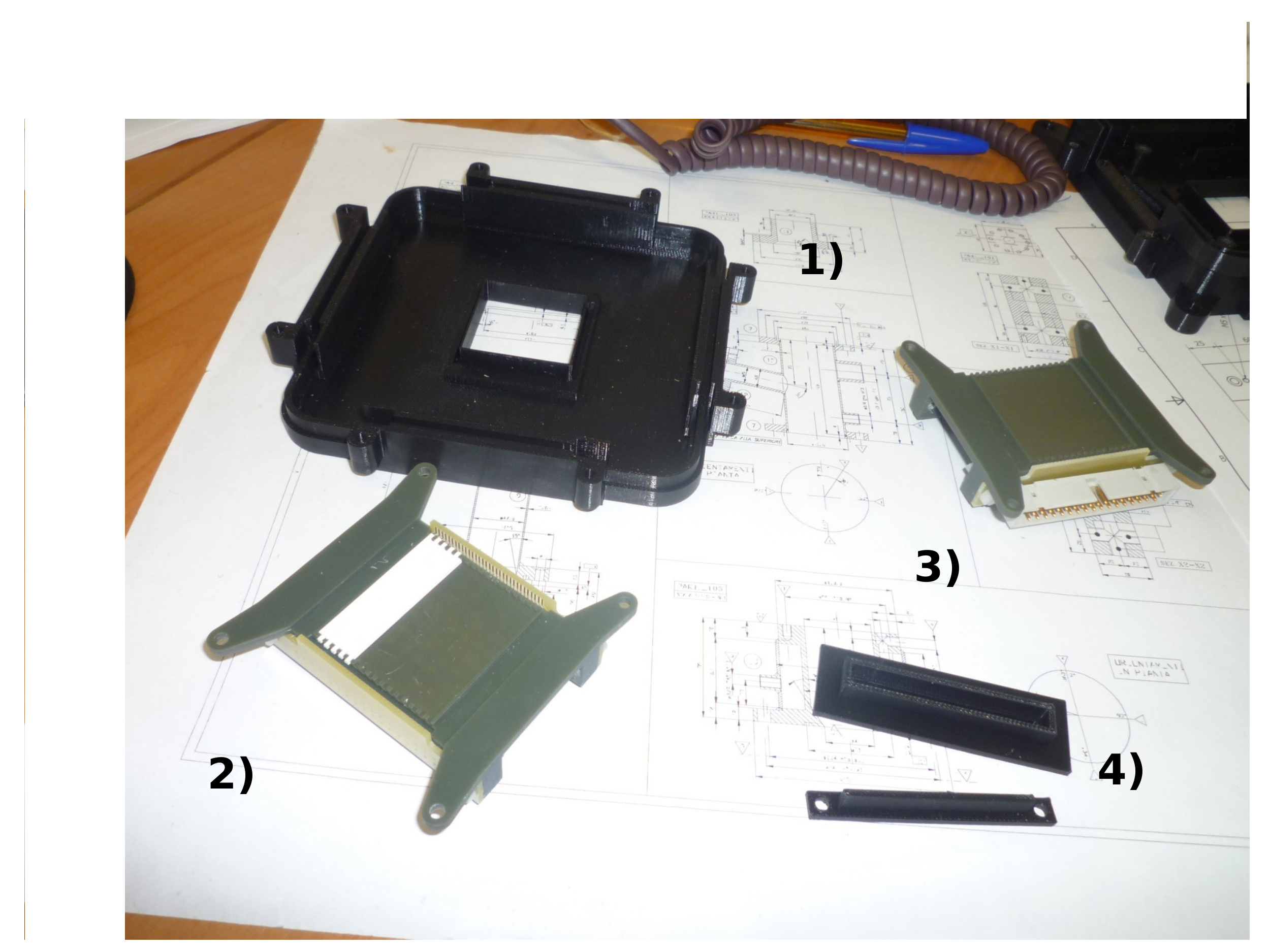}
\caption{Top panel: CAD design of the 1mm pitch hodoscope, with front and
         side views. Bottom  panel: main mechanical components of the detector.
         Item 1) is one of the two light-tight beam hodoscope cover boxes;
         item 2),3) are X/Y plane fiber holders: some fibers are mounted 
         in 2); 4) is the cable pass-through. }
\label{fib:1mm}
\end{figure}
The main requirements were to reduce to a minimum the material in front 
of the target thin Be window and to try to simplify the layout of the 
front-end electronics based 
        on one side on VME-like custom TPS boards, with a GPIB 
        to USB control module and on the other side on VME-compliant QDAC 
        and TDCs.
To solve the first issue, 32+32 1 mm$^2$ square scintillating Bicron BCF12 
fibers~\footnote{coated with white Extra Mural Absorber (EMA), to avoid light cross-talk} were 
used. They were arranged in two orthogonal planes along X/Y coordinates,
giving a detector fiducial area of $32 \times 32$ mm$^2$ .
As before, fibers were cut on a Fiberfin-4 machine at Cern, 
giving ready-to-use fibers with polished ends. 
The mechanics layout and the main mechanical components are shown in 
figure \ref{fib:1mm}.
The reduced size of mechanics details of some components gave severe
problems for the 3D-printing: issues were solved using an ENVISIONTEC
Perfactory  3D-printer at
LNGS with a precision better than 0.025 mm.
\begin{figure}[htbp]
\centering 
\includegraphics[width=.73\textwidth]{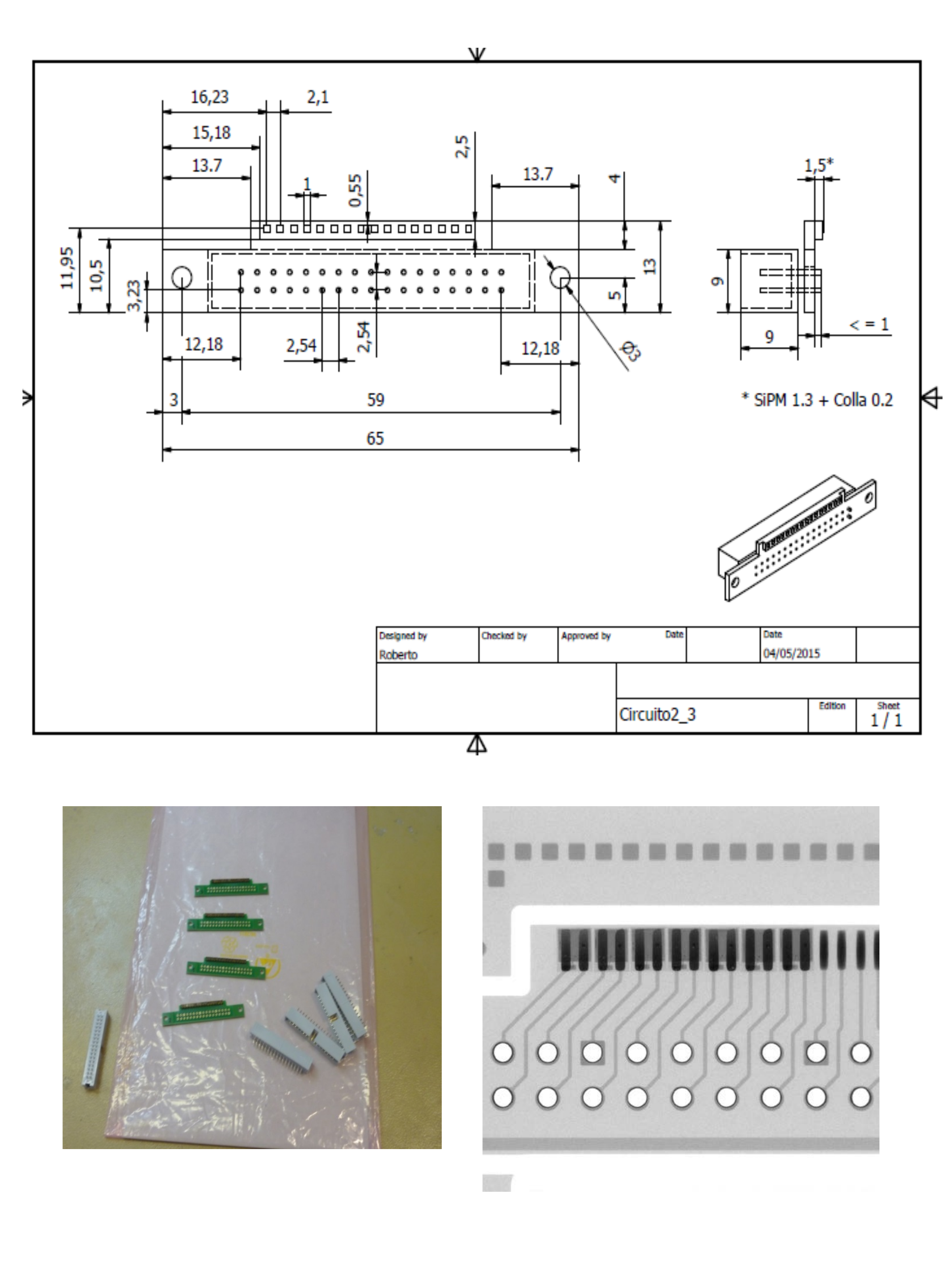}
\caption{Top panel: schematic layout of the PCB holding on one side the
16 SiPMs facing the fibers and on the other side the  34-way flat-cable
connector. This connector has a size 
comparable to the one of the PCB where it is mounted. 
Right  bottom panel: X-rays image to cross-check the SiPMs 
mounting on the PCB. Left bottom panel: picture of some PCBs and the 34-way
connectors. }
\label{fig:pcb1}
\end{figure}
$1 \times 1$ mm$^2$ square SiPMs from Advansid~\footnote{40$ \ \mu$m cells,
RGB type} were used to detect scintillation light emitted by crossing muons
in the detector's fibers. As the SiPM's footprint is slightly bigger than
the fiber cross-section, fibers had to be read alternating left/right and 
up/down sides, as for the R484 hodoscope. 
Each SiPM is mounted on a custom PCB (in groups of 16) 
as shown in figure \ref{fig:pcb1}. On the side facing the fibers the SiPMs
have been soldered~\footnote{The mounting was realized at Mevinco srl,
Soiano del Lago, Brescia and cross-checked with X-ray imaging, see the right
panel of figure \ref{fig:pcb1} for details}, while on the opposite side a 
34-way flat-cable connector was mounted, to convey bias and take out signals
from the SiPMs.  
Some steps of the mounting of the detector are shown in figure \ref{fig:ass2},
starting from the holder of one plane of fibers to the installation at
RIKEN-RAL.
\begin{figure}[htbp]
\centering 
\includegraphics[width=.8\textwidth]{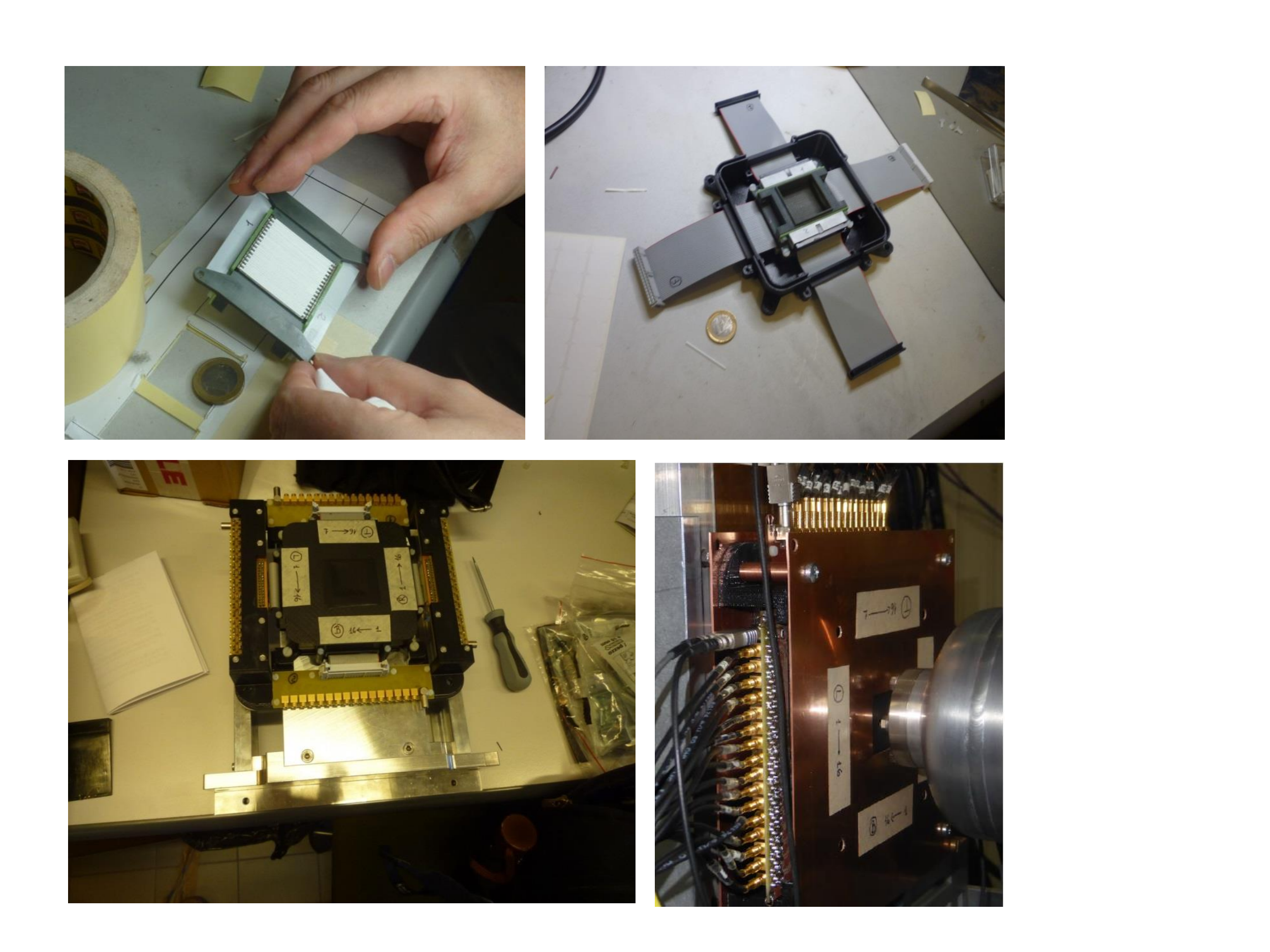}
\caption{Left-top panel: holder of one plane of fibers, with white EMA coated
fibers inserted. Right-top panel: mounted detector, without the cover box. 
Flat cables extend outside. Left-bottom panel: mounted detector, with an 
interface board visible on the bottom. Right-bottom panel: complete detector, with 1.5 mm copper planes
for electrical shielding installed at RIKEN-RAL. The cryogenic target nose
is visible on the right.}
\label{fig:ass2}
\end{figure}
Advansid RGB $1 \times 1$ mm$^2$ SiPMs were chosen as photodetectors for
the same reasons that dictated the choice for the $3 \times 3$ mm$^2$ 
SiPM of the R484 beam hodoscope. 
All available SiPMs (115 in all) were tested individually to determine
their breakdown voltages by measuring their current-voltage characteristic, 
see the left panel of figure \ref{fig:sipmt1} for an example.
\begin{figure}[htbp]
\vskip -2cm
\centering 
\includegraphics[width=.59\textwidth]{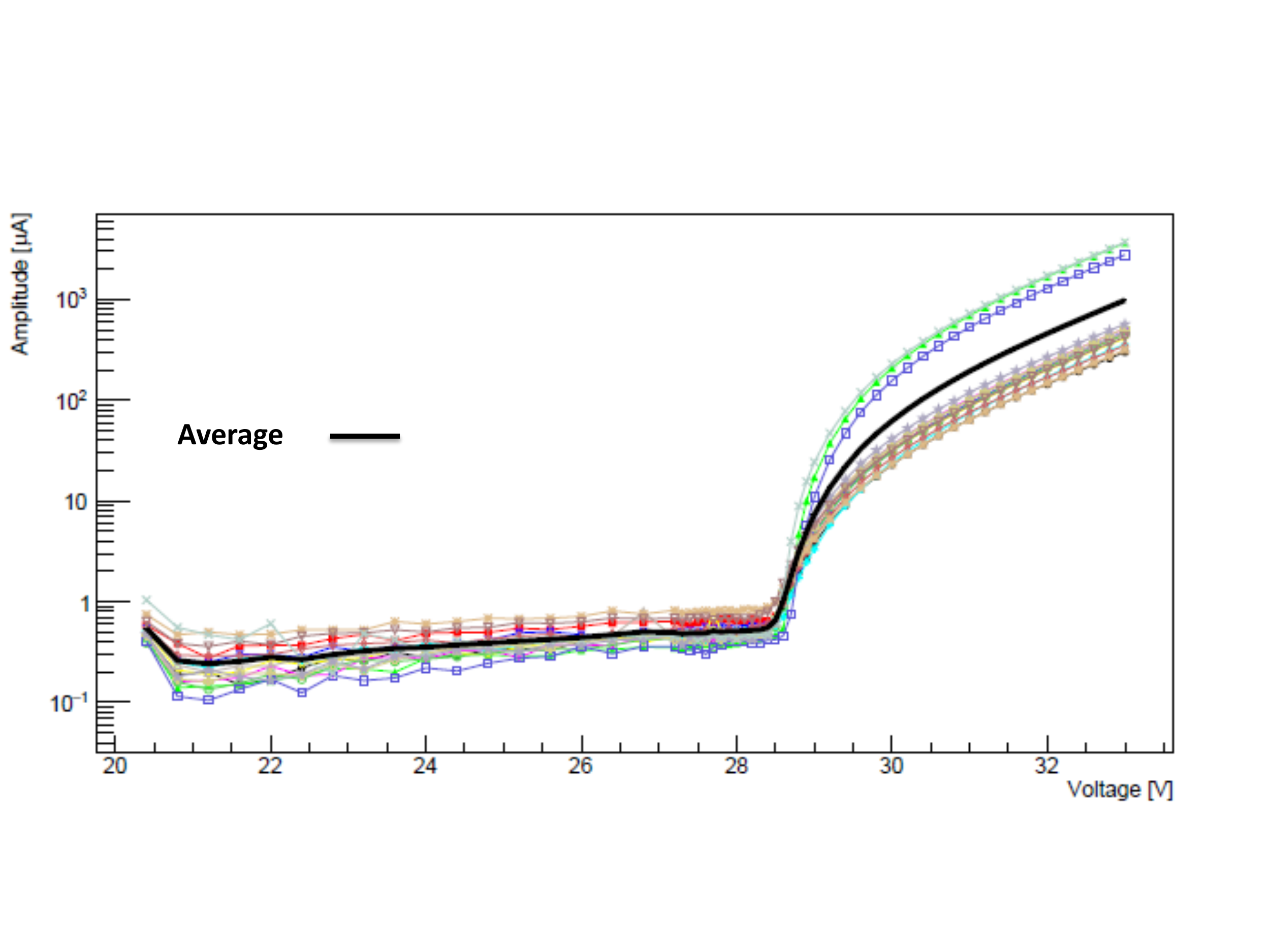}
\includegraphics[width=.39\textwidth]{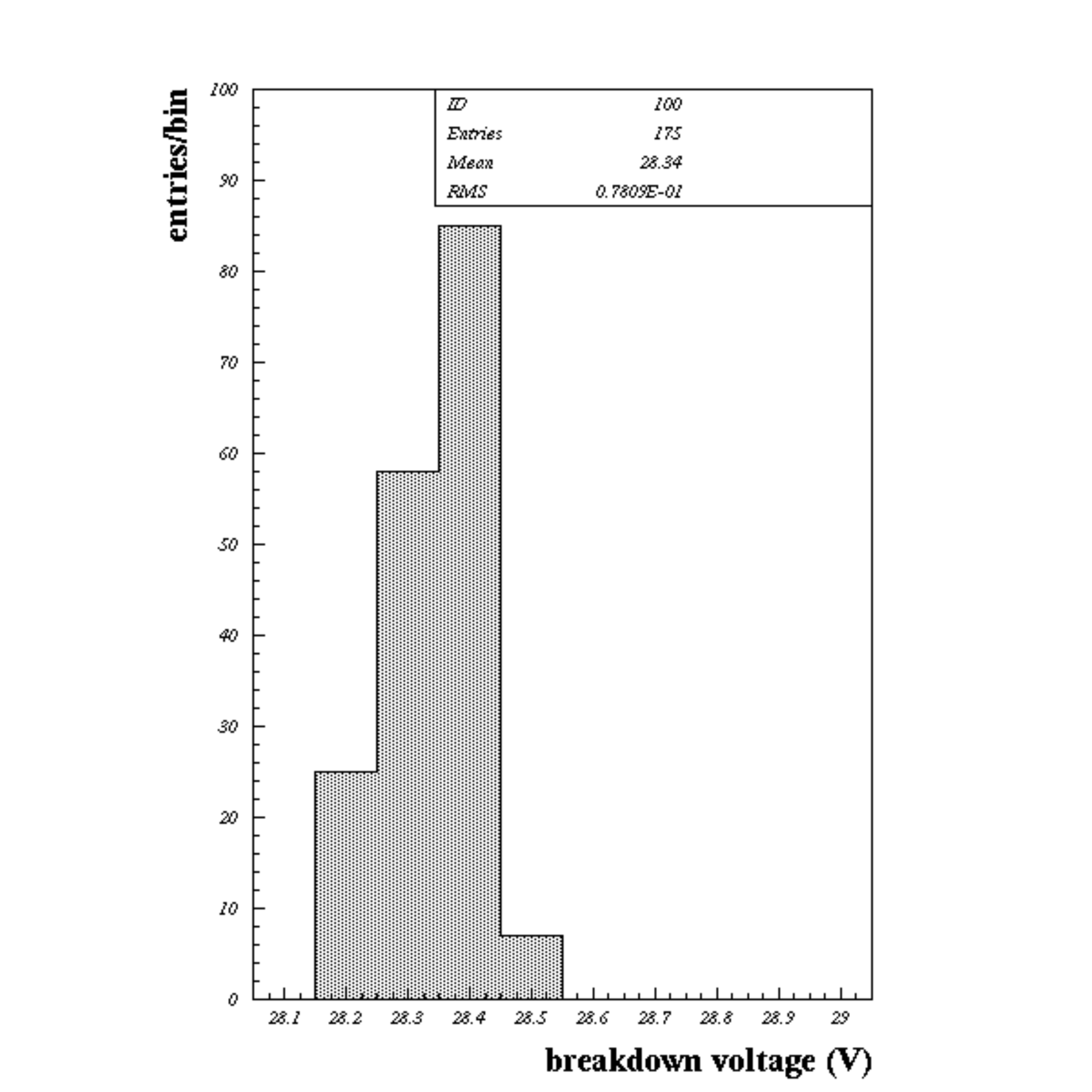}
\caption{Left panel: reverse I-V characteristic for the sample of 
$1 \times 1$ mm$^2$ SiPMs mounted on one side of the hodoscope. 
Right panel: distribution of $V_{brk}$ for the
sample of $1 \times 1$ mm$^2$ SiPMs. }
\label{fig:sipmt1}
\end{figure}
The right panel of figure \ref{fig:sipmt1} shows the distribution of their
breakdown voltages: on average 
$ 28.94 \pm 0.08 V$. It was thus possible, with a suitable
selection of the SiPMs to be used, to employ a common voltage for the 
biasing of the SiPMs of a detector plane. 
We checked that the 34-way flat cable, that had not a $50 \Omega$ impedance, 
gave no appreciable distortions to the signal, thanks to its short length.
Figure \ref{fig:board} shows one of the four interface boards, with 
the flat cable socket on one side and angled MCX connectors for 
signal cables and  one LEMO 00 connector for the power line on the other side.  
\begin{figure}[htbp]
\centering 
\includegraphics[width=\textwidth]{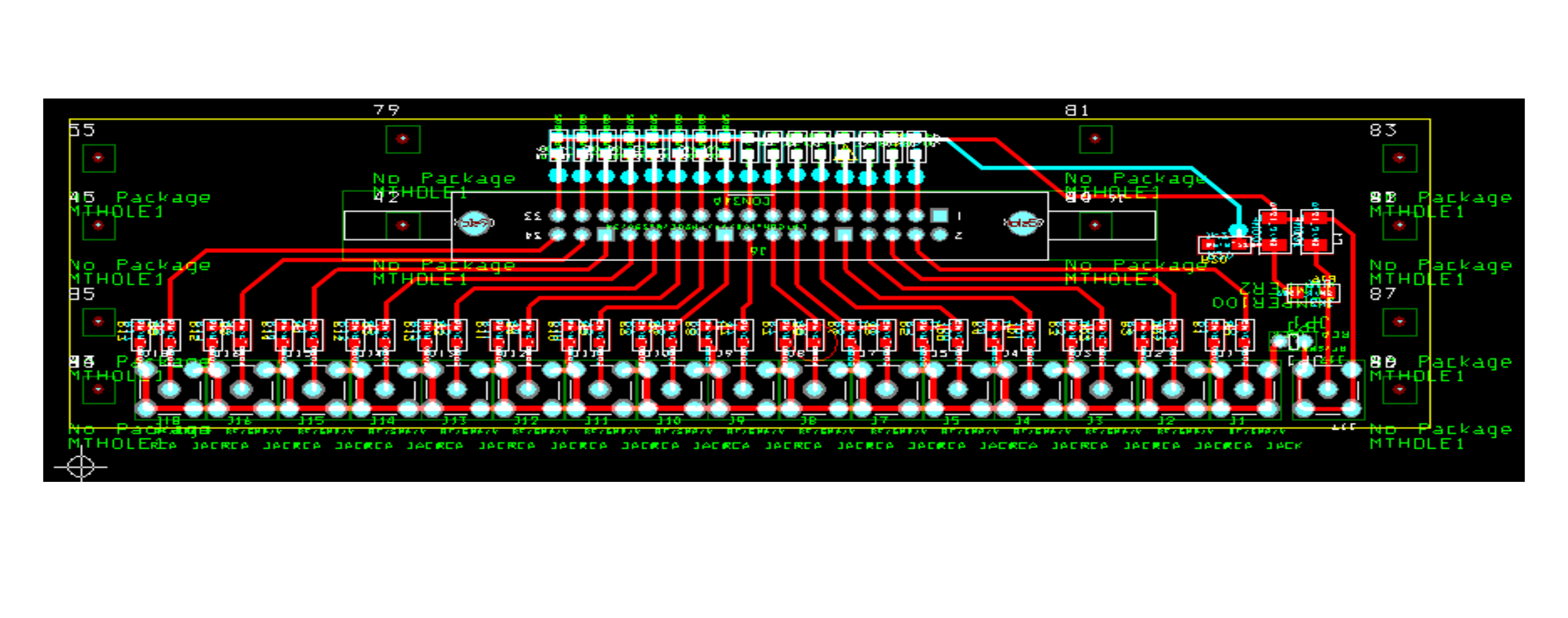}
\vskip -0.5 cm
\caption{Electronics rendering (GERBER) of the interface board for the beam
hodoscope of R584. The 34-way flat cable connector is placed on top.}
\label{fig:board}
\end{figure}
Signals are then fed into a fast FADC~\footnote{CAEN V1742 FAC with 5 GS/s,
12 bit, 1 Vpp input dynamic range,  in VME standard} operated at a reduced
sampling rate of 1 Gs/s to increase the digitizing buffer time.
 
Foreseen developments include the addition of onboard thermistors to ensure
long-term stability via correction of  SiPMs gain thermal drift and the
possibility to use also timing informations from the detector and the
monitor of the beam intensity.
The left panel of figure \ref{fig:res1} shows the
FADC waveform for a typical channel, where the two pulses beam structure is 
clearly visible.
Signals were then processed via a custom DAQ sysyem based on a CAEN V2718 
VME-PCI interface. 
\section{Preliminary performances at RIKEN-RAL.}
The two beam monitors have been used at RIKEN-RAL to optimize the
beam steering inside the target. For the R484 experiment, the integrated 
charge provided by the CAEN V792 QDC was directly used, while for
the R582 experiment  the signal waveform for each channel 
was integrated, after subtracting the baseline, providing the same type 
of infos. 
\begin{figure}[htbp]
\vskip -0.5cm
\centering 
\includegraphics[width=.7\textwidth]{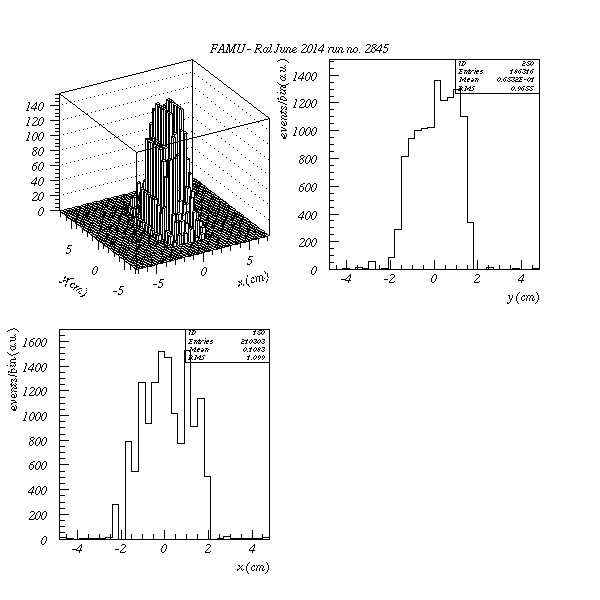}
\vskip -0.5 cm
\caption{X/Y beam profile for a nominal optics run in R484 at 61 MeV/c.}
\label{fig:res}
\end{figure}
Results for a typical run in R484 are shown in figure \ref{fig:res},
while results with standard optics for the R582 run are shown in figure
\ref{fig:res1}. In both cases the beam size is defined by the collimator 
aperture,  
with a measured RMS smaller than 10 mm.
\begin{figure}[htbp]
\centering 
\vskip -0.5cm
\includegraphics[width=.35\textwidth]{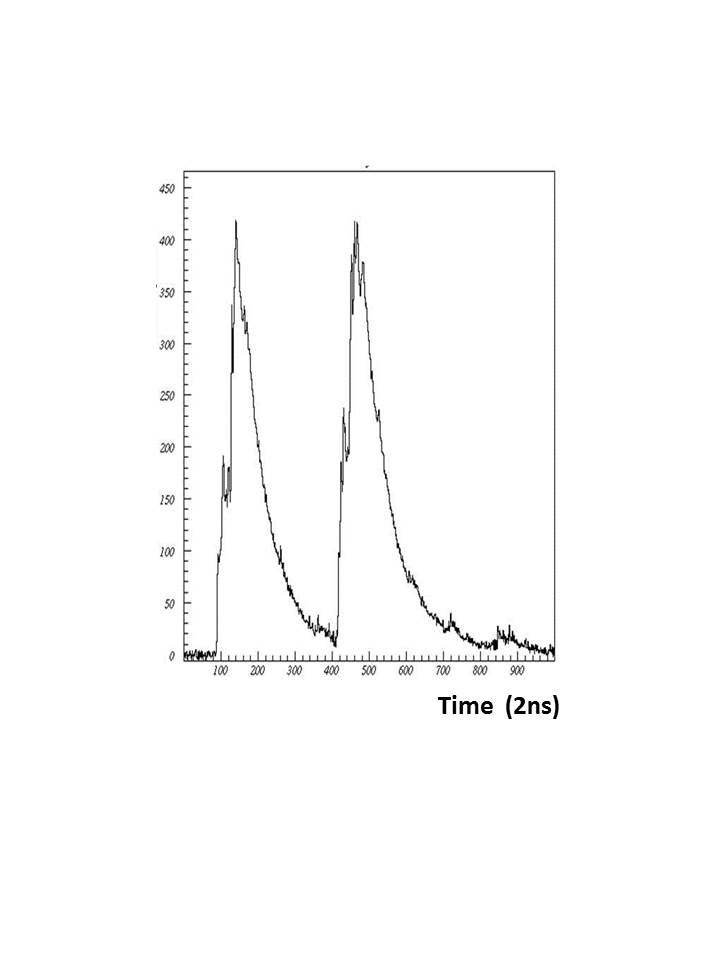}
\includegraphics[width=.64\textwidth]{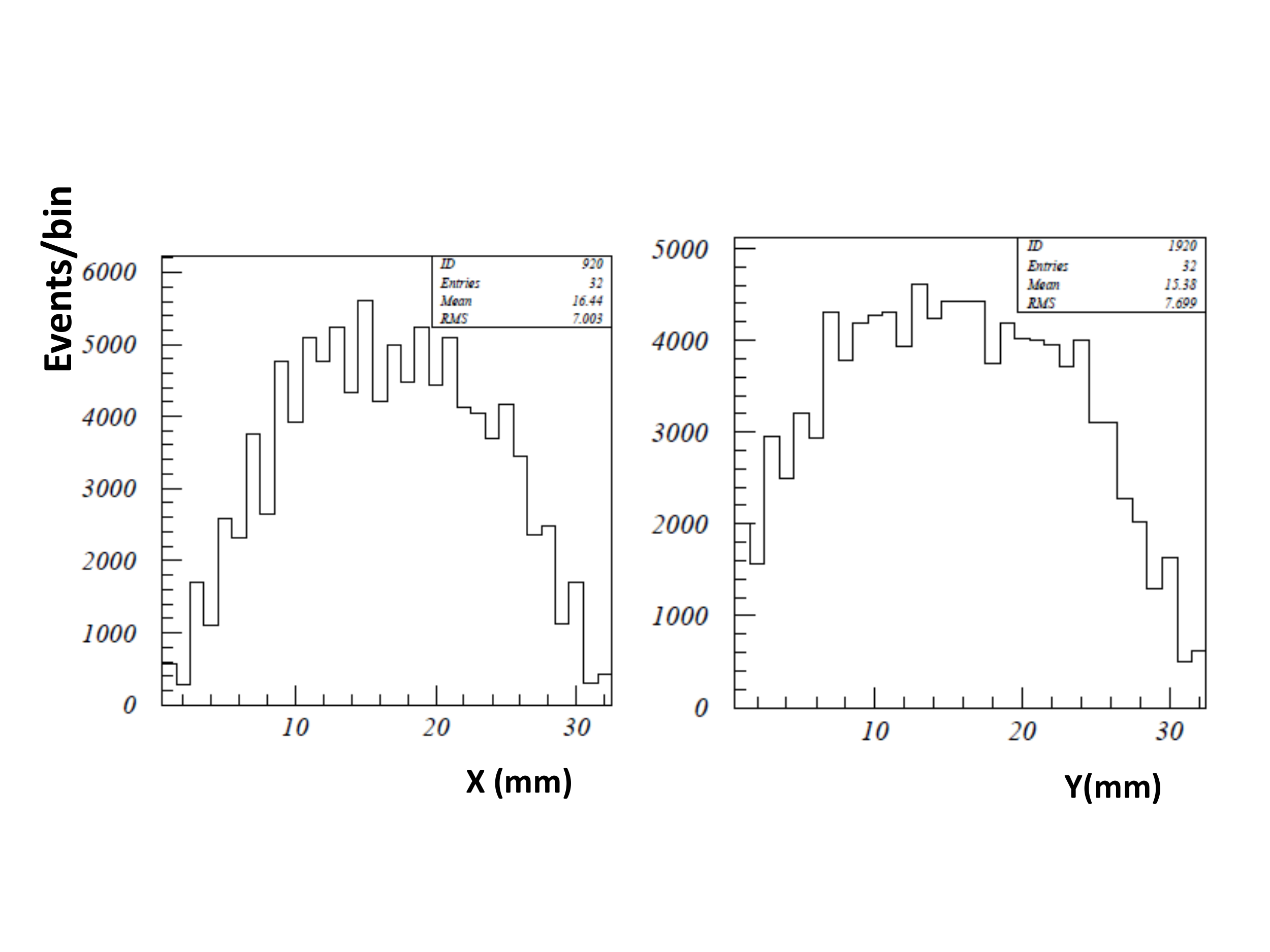}
\vskip -1.3cm
\caption{Left panel: FADC waveform for a typical beam hodoscope channel.
         Right panel: X/Y beam profile at RIKEN-RAL for a 60 MeV/c run.}
\label{fig:res1}
\end{figure}
\section{Conclusions}
\label{sec:conc}
Two scintillating fibers beam hodoscopes with 1 mm and 3 mm pitch and 
X/Y readout by
 SiPM have been developed and used for the measurement of 50-60 MeV/c 
beam profiles at RIKEN-RAL muon beams. Performances in agreement
with expectations are reported and foreseen developments include the
possibility to monitor the beam intensity and use timing informations.
\section*{Acknowledgements}
We would like to thank S. Banfi, M. Geigher (INFN Milano Bicocca), 
O.Barnaba,R.Nard\'o (INFN Pavia),
A. Iaciofano (INFN Roma Tre)
for help in mechanics and electronics. We acknowledge
the help  of T.Schneider (CERN) for fiber cutting at CERN,
D. Orlandi (LNGS) for 3D-printing of some items of the detector at LNGS,
N. Serra 
(Advansid) 
for 
SiPMs best handling and of P.Branchini and
D.Tagnani (INFN Roma Tre) for the optimal use of TPS electronics . 
We would like to thank also the staff of ISIS and the RIKEN-RAL facility for 
the generous help during data-taking and in particular  of K. Ishida for 
advice in the  optimal use of our detectors in beam. 

\end{document}